\documentclass[%
 reprint,
superscriptaddress,
showpacs,
 amsmath,amssymb,
 aps,
prd,
]{revtex4-1}

\usepackage{graphicx}
\usepackage{dcolumn}
\usepackage{bm}

\newcommand{\Cornell}{\affiliation{Cornell Center for Astrophysics and Planetary Science, Cornell University, Ithaca, New York, 14853, USA}}
\newcommand{\WSU}{\affiliation{Department of Physics \& Astronomy,
        Washington State University, Pullman, Washington 99164, USA}}
\newcommand{\CITA}{\affiliation{Canadian Institute for Theoretical
    Astrophysics, University of Toronto, Toronto, Ontario M5S 3H8, Canada}}

\newcommand{\TAPIR}{\affiliation{TAPIR, Walter Burke Institute for Theoretical Physics,
    California Institute of Technology, MC 350-17, Pasadena, CA 91125, USA}}

\newcommand{\BERKELEY}{\affiliation{Lawrence Berkeley National Laboratory, 1 Cyclotron Rd, Berkeley, CA 94720, USA}} %
\newcommand{\NEWHAMPSHIRE}{\affiliation{Department of Physics, University of New Hampshire, Durham, New Hampshire 03824, USA}} %
\newcommand{\AEI}{\affiliation{Max Planck Institute for Gravitational Physics (Albert Einstein Institute), D-14476 Potsdam-Golm, Germany}} %
\newcommand{\STOCKHOLM}{\affiliation{Department of Astronomy and Oskar Klein Centre, Stockholm University, AlbaNova, SE-10691 Stockholm}}

\usepackage{color}

\definecolor{cerise}{RGB}{222,49,99}

\begin{document}

\preprint{APS/123-QED}

\title{Black hole-neutron star mergers using a survey of \\ finite-temperature equations of state}

\author{Wyatt Brege}
\author{Matthew D. Duez} \WSU %
\author{Francois Foucart} \NEWHAMPSHIRE \BERKELEY
\author{M. Brett Deaton}
\author{Jesus Caro} \WSU %
\author{Daniel A. Hemberger} \TAPIR
\author{Lawrence E. Kidder} \Cornell
\author{Evan O'Connor} \STOCKHOLM
\author{Harald P. Pfeiffer} \AEI\CITA
\author{Mark A. Scheel} \TAPIR



\date{\today}

\begin{abstract}
Each of the potential signals from a black hole-neutron star merger should contain an imprint of the neutron star equation of state:  gravitational waves via its effect on tidal disruption, the kilonova via its effect on the ejecta, and the gamma ray burst via its effect on the remnant disk.  These effects have been studied by numerical simulations and quantified by semi-analytic formulae.  However, most of the simulations on which these formulae are based use equations of state without finite temperature and composition-dependent nuclear physics.  In this paper, we simulate black hole-neutron star mergers varying both the neutron star mass and the equation of state, using three finite-temperature nuclear models of varying stiffness.  Our simulations largely vindicate formulae for ejecta properties but do not find the expected dependence of disk mass on neutron star compaction. We track the early evolution of the accretion disk, largely driven by shocking and fallback inflow, and do find notable equation of state effects on the structure of this early-time, neutrino-bright disk.

\end{abstract}

\maketitle


\section{Introduction}

Compact neutron star binary mergers, whether composed of two neutron stars (NSNS) or of a neutron star and a black hole (BHNS) are strong gravitational wave sources and can produce counterparts across the electromagnetic spectrum.  Both signal types may contain imprints of the high-density equation of state (EOS).  The first observation of a NSNS merger, GW170817, demonstrated that NSNS binaries can produce at least low-energy short duration gamma ray bursts (GRBs)~\cite{TheLIGOScientific:2017qsa,GBM:2017lvd}.  A key difference between NSNS and BHNS systems is that NSNS mergers eject material away from the equatorial plane of the binary, while BHNS mergers do not.  A relativistic jet from an NSNS central remnant may break through this surrounding material or may be choked inside it; various scenarios of cocoon-jet interaction have been considered in models of GW170817/GRB170817A~\cite{Ioka:2017nzl,Mooley:2017enz,Margutti:2017cjl,Lazzati:2017zsj,Lyman:2018qjg}.  The production of standard short GRBs thus may proceed somewhat differently, and is perhaps easier, for BHNS mergers.  The strong EOS-dependence of the gravitational wave cutoff frequency~\cite{Vallisneri00,Pannarale:2015jia} and the post-merger disk and ejecta masses, making them conceivably EOS probes, are other attractive features of this system type.

\begin{figure}
\includegraphics[width=1.3\columnwidth]{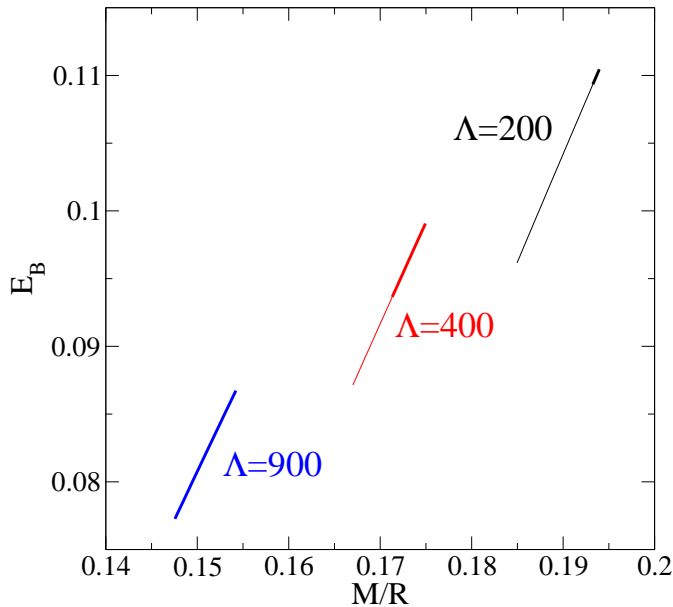}
\caption{Surfaces of constant tidal deformability $\Lambda$ for a two-component piecewise polytrope family, for which the pressure is $P=\kappa_0\rho_0^{\Gamma_0}$ for $\rho_0<\rho_T$, and $P=\kappa_1\rho_0^{\Gamma_1}$ for $\rho_0>\rho_T$.  A stellar model is specified by the EOS parameters 
$\Gamma_0$, $\kappa_0$, $\Gamma_1$, $\rho_T$ (with $\kappa_1$ given by continuity of $P$) plus the central density $\rho_c$.  The low-density EOS is known; we set $\kappa_0$, $\Gamma_0$ as in~\cite{Lackey:2013axa}.  We vary $\Lambda_1$ over the range 2.4--3.3 and solve for $\rho_T$ and $\rho_c$ to satisfy $M_{\rm NS}=1.35M_{\odot}$ and $\Lambda$ equal to its value on the contour.  Only the portions of contours with thick line width allow a neutron star with mass greater than 1.97$M_{\odot}$.}
\label{fig:isolambdas}
\end{figure}

Numerical relativity simulations have been used to fit analytic models for the gravitational waveform~\cite{Lackey:2013axa,Pannarale:2015jia}, the post-merger disk mass~\cite{Foucart2012}, and the mass and asymptotic speed of the dynamical ejecta~\cite{Kawaguchi:2016}.  In addition to dependencies on the black hole mass $M_{\rm BH}$ and spin $S_{\rm BH}$, and on the neutron star mass $M_{\rm NS}$ (by which in this paper we shall mean the ADM mass in isolation of the neutron star given its baryonic mass), EOS information enters into these formulae through their dependence on the tidal deformability $\Lambda$, the compaction $C=M_{\rm NS}/R_{\rm NS}$, and the binding energy $E_{\rm B} = (M_{\rm 0,NS}-M_{\rm NS})/M_{\rm 0,NS}$, where $R_{\rm NS}$ and $M_{\rm 0,NS}$ are the neutron star radius and baryonic rest mass, respectively.  This would seem to be a lot of information if all these fitted quantities could be connected to observables.  However, these three quantities are, although not completely degenerate, tightly related, as illustrated in Fig.~\ref{fig:isolambdas}.  There we show the variation of compaction and binding energy along contours of constant $\Lambda$ for a particular EOS family, the 2-component piecewise polytropes, with the high-density polytropic index covering the reasonable range $2.4<\Gamma_1<3$.  For a given $\Lambda$, $C$ will vary by about 5\%, $E_B$ by about 10\%.  The close connection between $\Lambda$ and $C$ for realistic neutron star models has been known for some time; the apsidal constant $k_2$ does not depend strongly on EOS~\cite{Hinderer2010}. That $E_B$ shows slightly more variation at a given $\Lambda$ is presumably why it is useful as a second parameter.

In addition, most previous studies in full GR have used simple EOS, most often polytropic or piecewise polytropic with Gamma-law thermal extensions to allow shock heating.  Piecewise-polytropes have the enormous advantages of fitting a wide range of barotropic EOS and allowing systematic variation of EOS parameters.  However, after the tidal disruption of the neutron star, the EOS is no longer one-dimensional:  the pressure $P$ is not only a function of baryonic density $\rho_0$, but also of composition, measured by the electron fraction $Y_e$, and, after shock heating, temperature $T$.  Continuing to assume a barotropic cold component (essentially, assuming that beta equilibrium will continue to hold) after disruption can potentially have unphysical effects~\cite{FoucartBhNs2016}, while the lack of physical temperature information makes it impossible to incorporate neutrino physics, which is crucial for the disk evolution and possibly for the production of GRBs.

Several numerical relativity studies have used nuclear-theory based $(\rho_0,Y_e,T)$-dependent EOS in tabulated form.  These include our previous simulations using the Shen~\cite{Shen:1998gq,Duez:2009yy}, Lattimer-Swesty~\cite{Lattimer:1991nc,Deaton2013,Foucart:2014nda}, and DD2 EOS~\cite{Hempel:2011mk,FoucartBhNs2016} and, most recently, Kyutoku~{\it et al}~\cite{Kyutoku:2017voj}.  The latter focus on a single set of binary parameters, with neutron star mass $M_{\rm NS}=1.35M_{\odot}$, aligned black hole spin 75\% the Kerr limit ($S_{\rm BH}/M_{\rm BH}^2=0.75$), and mass ratio 4:1, but use the DD2, SFHo, and TM1 EOS.

This paper extends these previous studies.  We simulate binary systems with a realistic black hole mass and black hole spin sufficient for strong electromagnetic counterparts.  The neutron star compaction depends both on the neutron star mass and the equation of state, so we vary both, looking for notable differences in the effect on merger observables.

We observe the effects noted in earlier studies of neutron star mass and compaction on the dynamical ejecta, and we find that more compact stars tend to produce more compact, more neutrino luminous early-time disks.  
We compare gravitational wave, ejecta, and disk properties with analytic predictions based on simulations with less realistic EOS.  For the most part, we confirm the validity, within expected errors, of these formulae.  However, disk mass does not decrease as expected with increased compaction in this region of parameter space.  This is, perhaps, an indication that disk mass is less sensitive to compaction for binary systems that produce large disk masses.  Finally, we present a detailed analysis of the three major components of the post-merger matter distribution:  the ejecta, the incipient accretion disk, and the fallback material.  

This paper is arranged as follows.  In Section~\ref{sec:eos}, we discuss our numerical methodology and the equations of state employed.  In Section~\ref{sec:results}, we present results for the post-merger outputs.  In Section~\ref{sec:discussion}, we discuss the future evolution of the system and gather conclusions on EOS signatures in BHNS mergers.

\begin{figure}
\includegraphics[width=1.3\columnwidth]{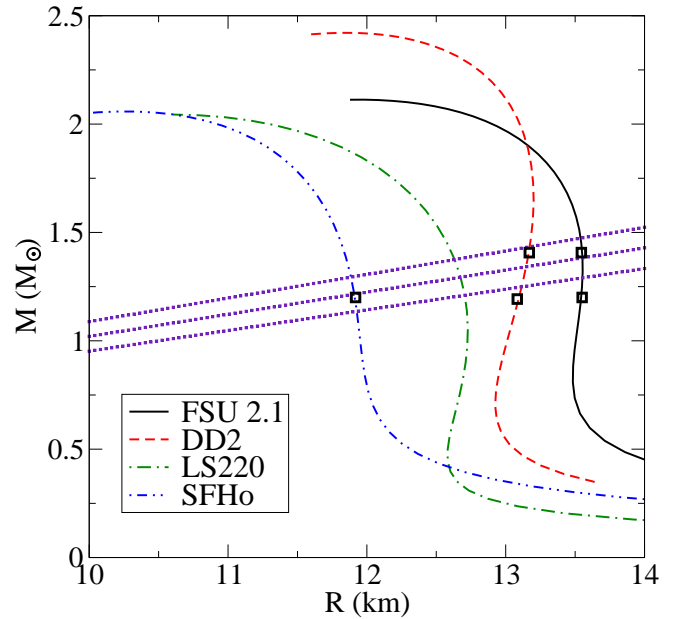}
\caption{ADM neutron star mass vs. areal radius for nuclear equations of state sliced at $T=0.1{\rm MeV}$ in $\beta$-equilibrium. Black boxes mark the stars used for this survey, which are chosen to have ADM masses 
$M_{\rm NS} = (1.2, 1.4) M_{\odot}$.  The three dotted indigo curves are contours of constant compaction.  From botton to top, they are $C=$0.14, 0.15, and 0.16.}
\label{fig:tov}
\end{figure}

\section{Equations of State and Binary Models}
\label{sec:eos}

We use three finite-temperature, composition-dependent nuclear-theory based equations of state, all based on relativistic mean field models (RMFs) and publicly available in tabulated form at \url{http://www.stellarcollapse.org}~\cite{OConnor2010}.
\begin{enumerate}
\item {\bf FSUGold}~\cite{todd2005neutron,Shen:2011kr,shen2011second}:  a RMF with modifications at high density to increase the maximum neutron star mass to 2.1\,$M_{\odot}$.  This EOS predicts a radius of $R_{\rm NS}=$13.5\,km and tidal deformability of around $\Lambda=$970 for a 1.35\,$M_{\odot}$ neutron star.
\item {\bf DD2}~\cite{typel2010composition,Hempel:2011mk}:
another RMF with a density-dependent nucleon-meson coupling, giving $R_{\rm NS}= $13.1\,km, $\Lambda=860$ for a 1.35\,$M_{\odot}$ neutron star.
\item {\bf SFHo}~\cite{steiner2013core}: an RMF using a covariant Walecka model Lagrangian (ensuring causal sound speeds) with parameters specifically designed to match most-probable neutron star properties as inferred by observations~\cite{2013ApJ...765L...5S}.  This means more compact stars: SFHo gives $R_{\rm NS}=$11.8\,km and $\Lambda=420$ for a 1.35\,$M_{\odot}$ neutron star.  We also attempted simulations with the even-softer SFHx EOS, but numerical errors during tidal disruption proved too large for simulations to continue to completion.
\end{enumerate}

$M$-vs-$R$ curves for these EOS (evaluated at low temperature and neutrinoless beta-equilibrium $Y_e$) are plotted in Fig.~\ref{fig:tov}.  Many of our previous BHNS simulations~\cite{Deaton2013,Foucart:2014nda} used the Lattimer-Swesty EOS with incompressibility $K=220$MeV~\cite{Lattimer:1991nc}, which is also included in Fig.~\ref{fig:tov} for comparison.  Unfortunately, the LS220 runs used an older version of SpEC without adaptive fluid grids and are insufficiently accurate to be included in the quantitative comparisons below.

All our binaries use a 7\,$M_{\odot}$ black hole, slightly below the peak of the distribution of observed black hole masses in X-ray binaries~\cite{Kreidberg:2012}.  The black hole spins in a prograde direction at 90\% of its Kerr limit.  Using $S_{\rm BH}/M_{\rm BH}{}^2=0.9$ allows us to compare with our previous DD2 studies~\cite{FoucartBhNs2016}, which used this spin, and also to explore a case not yet covered by Kyutoku~{\it et al}~\cite{Kyutoku:2017voj}.  For each EOS, we evolve a binary with a 1.2\,$M_{\odot}$ neutron star and a binary with a 1.4\,$M_{\odot}$ neutron star.  Unfortunately, the highest-compaction case, a 1.4$M_{\odot}$ neutron star with SFHo EOS, had unacceptably large evolution errors, and its simulation could not be completed.  This leaves five cases.  The neutron star fluid is taken to be irrotational.  At the chosen initial separation, the binaries proceed for about 5 orbits before merger.  The orbital eccentricity is of order $0.03-0.04$.

We evolve using the SpEC code.  Details of SpEC's methodology for non-vacuum systems can be found in our earlier papers~\cite{Foucart:2013a,FoucartBhNs2016}.  To summarize, the spacetime is evolved pseudospectrally on one grid, while the fluid is evolved using conservative shock-capturing techniques on another grid.  We use our new adaptive mesh technology for the fluid grid, described in a recent paper~\cite{FoucartBhNs2016}.  It combines higher resolution near the black hole with an ability to place grid boxes only in the proximity of matter.  Neutrino effects are treated using a 3-flavor energy-integrated neutrino leakage scheme, which can capture effects on the fluid of emitting neutrinos but not of absorbing them~\cite{Deaton2013,Foucart:2014nda}.

During inspiral, our standard resolution for the fluid grid covering the neutron star has grid spacing $\Delta x\approx 200$m. 
During merger, the fluid grid allows up to 7 nested layers of grid boxes; $\Delta x$ doubles with each layer outward.  The innermost box -- centered on the black hole -- covers a half-width of around 40\,km with $\Delta x\approx 240$m.
Our previous study~\cite{FoucartBhNs2016} reports convergence tests for BHNS binaries using the DD2 EOS and resolutions similar to ours.  We have also simulated the plunge and early merger phase (about 4\,ms) of two cases in the current study at 20\% lower resolution:  FSU with a 1.4\,$M_{\odot}$ star and SFHo with a 1.2\,$M_{\odot}$ star.  We find that post-merger mass predictions agree to 10\% for unbound matter and to 1\% for total baryonic mass outside the black hole (with more ejecta at higher resolution), while the ejecta average velocity and black hole irreducible mass track each other almost identically.  Assuming second order convergence, this would correspond to 20\% and 2\% errors in ejecta and disk mass, respectively.  This would be in addition to any errors related to initial data and inspiral, the former being difficult to assess because our usual eccentricity reduction procedure was not very effective for the small initial binary separations used in this study.  Resolutions of the sort used here are needed to track the thin stream of matter that flows to the black hole when the neutron star tidally disrupts.  If a segment of this stream is less than about 10 points across, unphysical heating, shocks, and mass ejection can result.  We check for the absence of such symptoms in simulations at the resolutions reported here.

\section{Results}
\label{sec:results}

Qualitatively, all mergers proceed in the same way.  The binary components inspiral due to gravitational radiation until the neutron star tidally disrupts.  The outer regions of the neutron star accelerate outward to become the dynamical ejecta.  Lagrangian tracer particles in this region show that, in the coordinates of our simulation, the orbital energy $e\equiv -u_t-1$ of this material begins negative but grows primarily due to gravitational torques and asymptotes by 1\,ms after disruption at positive values.  Meanwhile, inflowing matter forms a thin stream curving into the black hole.  Resolving the width of this stream well enough to avoid unphysical shocks was the primary computational challenge of this project.  Eventually, the stream intersects and shocks itself, forming a hot, roughly axisymmetric proto-disk.  This proto-disk is surrounded by infalling cold matter.  Material is still falling back and accumulating onto the proto-disk at a rapid rate 10\,ms later, when we terminate our simulations.  The subsequent evolution of the system will be discussed in Section~\ref{sec:discussion}.

\begin{table*}
\begin{center}
\begin{tabular}{l ccccccccccc }
\toprule \toprule
EOS & $M_{\rm NS}\,(M_\odot)$ & $R_{\rm NS}\,({\rm km})$ & $\cal{C}_{\rm NS}$ & $N_{\rm orbits}$ & $\Omega_0 M$ & $M_{\rm BH}^f\,(M_\odot)$ & $\chi_{\rm BH}^f$ & $M_{\rm out}^f\,(10^{-2} M_\odot)$ & $M_{\rm ej}\,(10^{-2} M_\odot)$ & $\langle v/c \rangle_{\rm ej}$ & $f_{\rm cut}M$\\
        \hline
        DD2 & 1.2 & 13.1 & 0.135 & 5.0 & 0.0426 & 7.7 [7.7] & 0.92 [0.93] & 37 [36] & 7.2 [7.2] & 0.21 [0.22] & 0.055\\
        DD2 & 1.4 & 13.2 & 0.156 & 6.0 & 0.0437 & 7.8 [7.9] & 0.92 [0.93] & 41 [34] & 6.0 [3.6] & 0.20 [0.21] & 0.07\\
        FSU21 & 1.2 & 13.5 & 0.130 & 4.9 & 0.0489 & 7.7 [7.7] & 0.92 [0.93] & 39 [38] & 7.9 [10.7] & 0.20 [0.22] & 0.051\\
        FSU21 & 1.4 & 13.6 & 0.152 & 5.5 & 0.0437 & 7.8 [7.9] & 0.92 [0.93] & 40 [36] & 5.9 [6.4] & 0.19 [0.21] & 0.068\\
        SFHo & 1.2 & 11.9 & 0.148 & 5.3 & 0.0489 & 7.7 [7.8] & 0.91 [0.93] & 37 [30] & 4.1 [4.3] & 0.18 [0.21] & 0.072\\
\hline \hline
\end{tabular}
\caption[Initial parameters of binaries used in this survey]{
Initial and final parameters of the binaries studied in this work.  Bracketed numbers are the predictions of analytic relations fit to prior simulations.  $M_{\rm NS}$ is the ADM mass of an isolated neutron star with the same equation of state and baryon mass as the neutron star under consideration, $N_{\rm orbits}$ is the number of orbits up to the point at which $0.01M_\odot$ has been accreted by the black hole, $\Omega_0$ is the initial angular velocity, and the system mass is $M=M_{\rm BH}+M_{\rm NS}$.  $M_{\rm BH}^f$ and  $\chi_{\rm BH}^f$  are the mass and dimensionless spin of the black hole, and $M_{\rm out}^f$ is the baryon mass remaining outside of the black hole.   The baryon mass outside the black hole is measured 10\,ms after merger.
$M_{\rm ej}$ is the mass of the dynamical ejecta, and $\langle v/c\rangle_{\rm ej}$ its mass-weighted average velocity. These properties are nearly constant, from about $1\,{\rm ms}$ after the merger. Bracketed numbers for $M_{\rm out}^f$ and $M_{\rm ej}$ show semi-analytical predictions for the mass outside of the black hole $10\,{\rm ms}$ after merger~\cite{Foucart2012}, and the ejected mass~\cite{Kawaguchi:2016}, while bracketed numbers for $M_{\rm BH}^f$ and $\chi_{\rm BH}^f$ are semi-analytical predictions from~\cite{Pannarale:2014}.  $f_{\rm cut}$ is the frequency at which the gravitational wave spectrum $fh(f)$ has dropped by a factor of two from its plateau (cf.~\cite{Kyutoku:2011vz}). 
}
        \label{tab:id}
        \end{center}
\end{table*}

The quantitative outcomes of the mergers are summarized in Table~\ref{tab:id}, which reports the final mass and spin of the black hole, the remaining mass of bound matter, the mass and asymptotic speed of the unbound ejecta, and the gravitational wave cutoff frequency, defined similarly to the definition in~\cite{Kyutoku:2011vz}.  These quantities can be compared to predictions derived from earlier simulations without finite-temperature nuclear EOS, or from analytic fits to those simulations.  Analytic formulae are available for bound mass 10\,ms after merger~\cite{Foucart2012}, for the post-merger black hole properties~\cite{Pannarale:2014}, and the ejecta properties~\cite{Kawaguchi:2016}.  For the ejecta velocity, a correction must be applied to account for the fact that our simulations roughly advect $Y_e$ while the EOS used for~\cite{Kawaguchi:2016} effectively enforce instantaneous beta equilibrium; the correction is described in~\cite{FoucartBhNs2016}.   These predictions are included in Table~\ref{tab:id} in brackets.

Overall, the agreement is within expected ranges.  This agreement is a nontrivial finding, given the EOS physics neglected in the simulations used to calibrate the formulae.  There are a couple of notable differences, however.  The ejecta velocity in these simulations is always slightly lower than the expected value, even with corrections for the different $Y_e$ evolution.  The mass outside the black hole matches the analytic ``disk mass'' prediction well for cases with low-compaction stars (DD2 and FSU21 with $M_{\rm NS}=1.2M_{\odot}$)  but is somewhat above the predicted values for the more compact stars. 

More interestingly, the expected pattern that more compact neutron stars should lead to less massive disks is not seen.  Compaction effects can be seen by comparing the same EOS at different $M_{\rm NS}$ or comparing different EOS for the same $M_{\rm NS}$.  In the former comparison, binaries with more massive and compact stars have slightly more massive disks.  This is also true at earlier times (e.g.$\sim 5$\,ms after merger).  In the latter comparison, the merger with SFHo produces a disk with roughly the same mass as that produced using DD2 or FSU2.1, even though SFHo yields a significantly more compact neutron star.  For comparison, Kyutoku~{\it et al} found an SFHo disk mass about $\frac{2}{3}$ that of DD2 for the slightly less extreme mass ratio 4 and black hole spin $S_{\rm BH}/M_{\rm BH}^2=0.75$.  (See Fig. 5 of~\cite{Kyutoku:2017voj}.)

Due to our grid spacing's scaling with $M_{\rm NS}$
, numerical evolution error is probably slightly higher in the simulations with more compact stars, but the convergence tests (which both involve these stars) suggest these errors are not large enough to explain the effect.  Error in initial conditions, as evidenced by the roughly 3\% initial orbital eccentricity in most simulations, may also contribute to error in disk masses, but we found no sign of systematically higher initial data error in more compact cases. The differences in disk mass between cases in Table~\ref{tab:id} is probably within numerical errors, but our accuracy is sufficient to suggest a softening of the connection between disk mass and compaction in the high black hole spin or high disk mass regime.  From our previous studies of BHNS mergers with neutron stars modeled as $\Gamma=2$ polytropes, it would appear that high disk mass is the deciding factor.  In an earlier work, we varied the star's compaction while setting $S_{\rm BH}/M_{\rm BH}^2=0.9$ with a higher mass ratio of 7, yielding lower disk masses, and the expected sensitivity of disk mass to compaction was seen~\cite{Foucart:2013a}.  

We should note that the analytical formula for the remnant disk mass~\cite{Foucart2012} is nominally valid only for smaller disk masses ($\lesssim 0.2M_\odot$), and typically underestimates disk masses for $0.2M_\odot \lesssim M_{\rm out}^f$, as seen both in SpEC simulations~\cite{Lovelace:2013vma} and by Kyutoku~{\it et al}~\cite{Kyutoku:2017voj}. That disk masses are higher than predicted for the more compact stars is thus less surprising than the good agreement observed for less compact stars. This agreement may be serendipitous. What is notable is that we do not produce higher disk masses for larger neutron stars, and that all disk masses measured in our simulations are within a very small range $0.37M_\odot < M_{\rm out}^f < 0.41M_\odot$, despite the use of very different equations of state in the simulation.

The cutoff frequency is easily consistent with Figure~24 of~\cite{Kyutoku:2011vz} (though this study did not include black hole spins quite as high as ours).  Most of the gravitational wave signal is in the (2,$\pm$2) modes.  The next-highest modes, (3,$\pm$3), and (4,$\pm$4), cut off at the same time as the dominant (2,$\pm$2) modes.

\begin{figure}
\includegraphics[width=1.2\columnwidth]{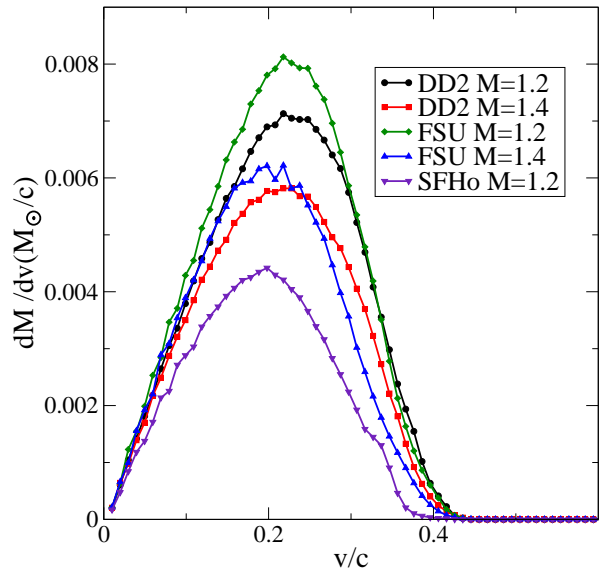}
\caption{Distribution of the asymptotic velocity of the ejecta measured 5 ms after merger.}
\label{fig:ejecta_asymptotic_velocity}
\end{figure}

\begin{figure}
\includegraphics[width=1.2\columnwidth]{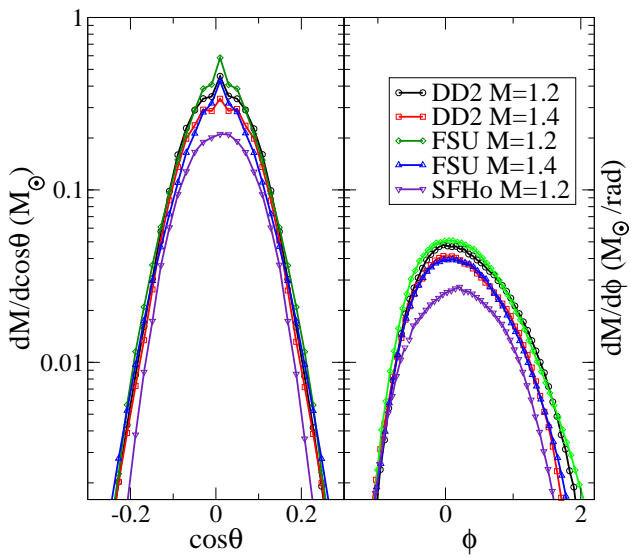}
\caption{Angular distribution in both $\theta$ and $\phi$ of the ejecta 5 ms after merger.  Most of the ejecta matter is constrained around the equator, within $\Delta \theta \sim 0.1$ radian.  In $\phi$, ejecta spans approximately half of the zonal sky, with an angle of $\Delta \phi \sim \pi$
, where the $M_{\rm NS}=1.2M_{\odot}$ SFHo case has the smallest arc and the $M_{\rm NS}=1.2M_{\odot}$ FSU2.1 case has the widest. }
\label{fig:ejecta_angular_distribution}
\end{figure}

\begin{figure}
\includegraphics[width=1.2\columnwidth]{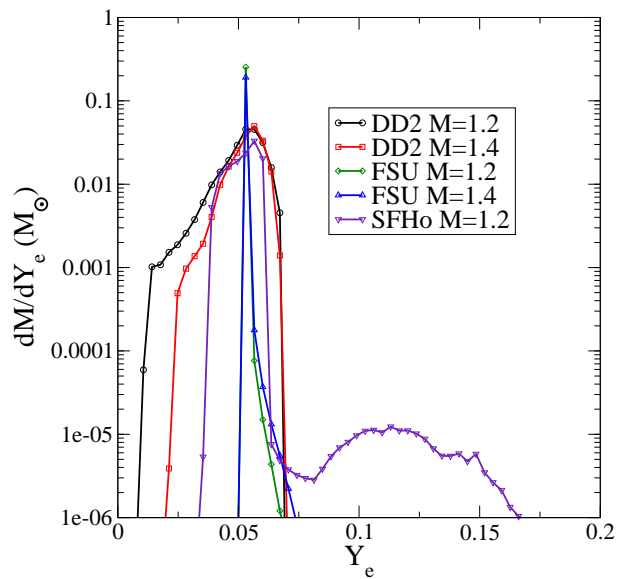}
\caption{Electron fraction $Y_e$ of the ejecta measured 5 ms after merger.  We note that all of the matter peaks in the $Y_e \approx 0.05$ range, where $M_{\rm NS}=1.2M_{\odot}$ DD2 has the largest electron fraction range ($0.011 - 0.07$).  The SFHo simulation has a distinct tail of $Y_e$ extending to around 0.2, but it has extremely little mass.  That neither FSU2.1 models produce ejecta with $Y_e < 0.05$ is an artifact of the bounds of the FSU2.1 table, which does not allow for $Y_e < 0.05$.}
\label{fig:ejecta_Ye}
\end{figure}

Ejecta properties are reported in more detail in Figures~\ref{fig:ejecta_asymptotic_velocity}, \ref{fig:ejecta_angular_distribution}, and \ref{fig:ejecta_Ye}.  In all cases, the asymptotic speed is around 0.2c, with a spread of $\approx 0.2$c above and below this.  In direction, the outflow is concentrated near the equator but fills an arc of about $\pi$ radian in the azimuthal direction, all consistent with previous studies~\cite{Kyutoku:2013,FoucartBhNs2016}. No dependence of this angular distribution on the EOS is apparent.

A prior study of BHNS dynamical ejecta with neutrino transport found that neutrino absorption has a negligible effect on ejecta~\cite{Kyutoku:2017voj},
which remains neutron-rich and should robustly produce 2nd and 3rd-peak r-process elements.  The unimportance of neutrino absorption gives us some confidence in the validity of our neutrino leakage results, at least as applied to the ejecta.  Our study also finds that the ejecta maintains low $Y_e$, as shown in Figure~\ref{fig:ejecta_Ye}.  There is a small ``bump'' at higher $Y_e$ for the soft SFHo EOS, but it has very little mass and still has $Y_e<0.2$.  Material at these low $Y_e$ will produce the second and third but not the first peak r-process elements~\cite{Lippuner2015}.

\begin{figure*}
\includegraphics[width=1\columnwidth]{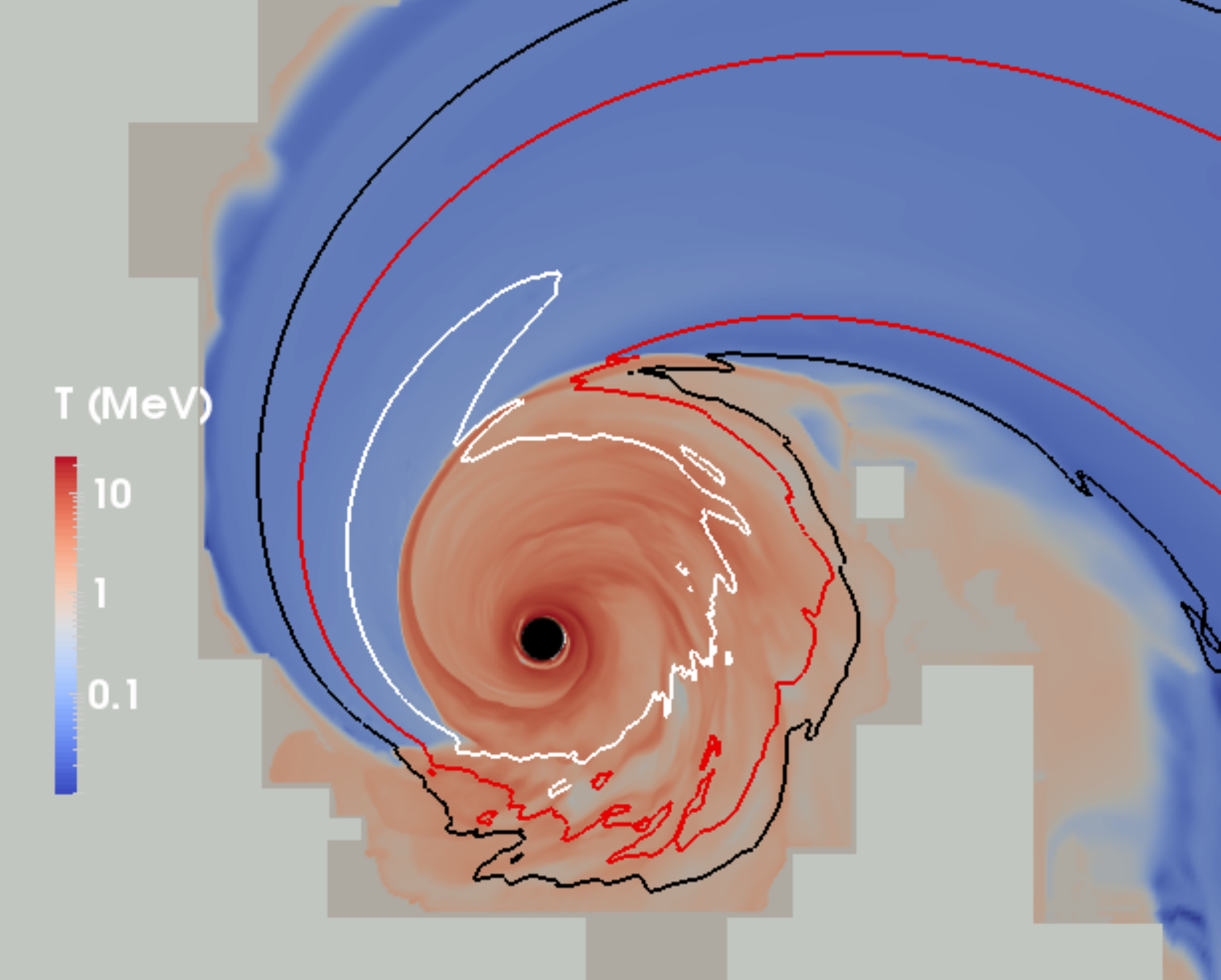}
\includegraphics[width=1\columnwidth]{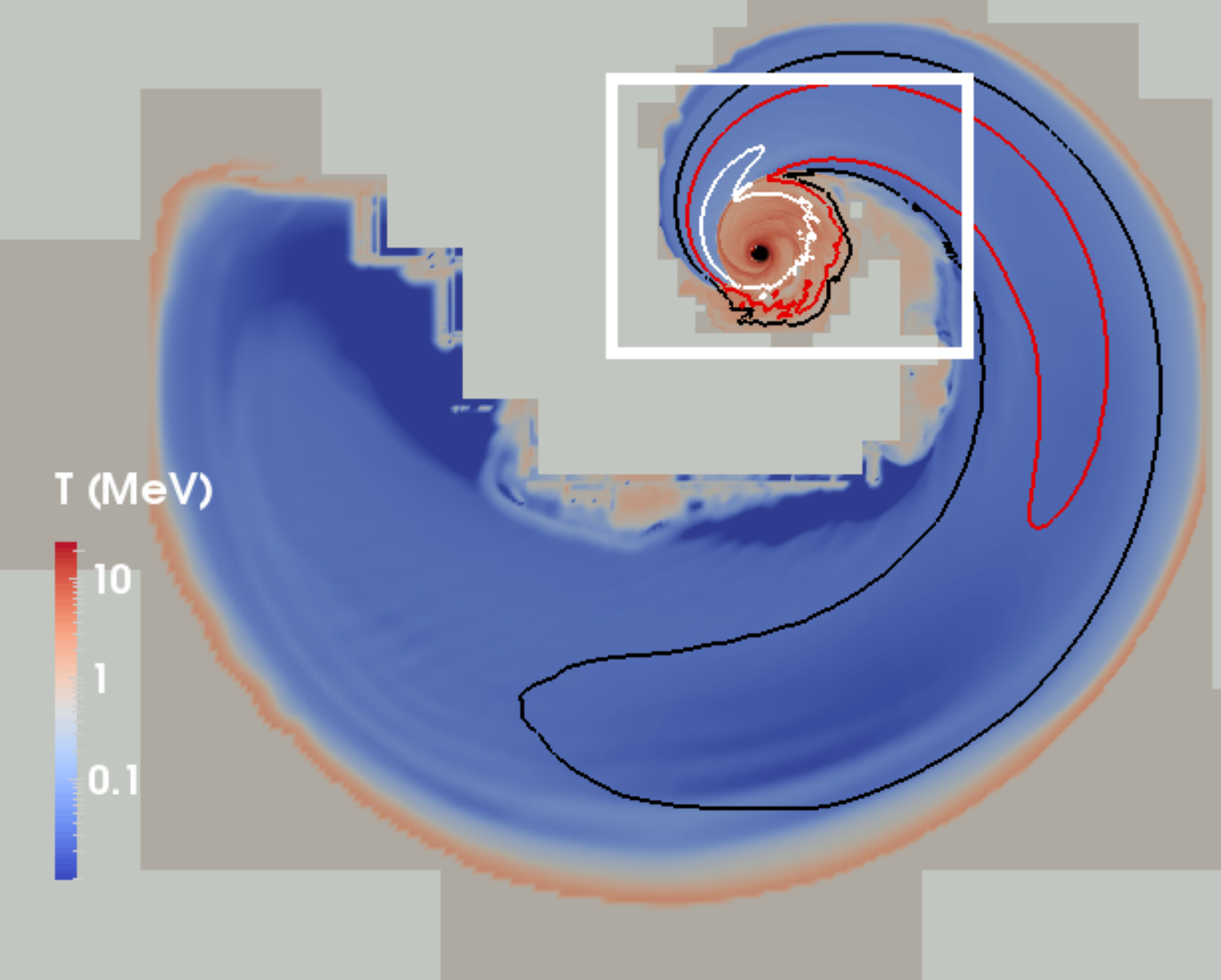}
\caption{Equatorial snapshot 7\,ms after merger of the FSU2.1 $M_{\rm NS}=1.2M_{\odot}$ case.  Left shows the black hole, the hot  accretion disk, and the inner part of the tidal tail.  The right image zooms out to show the entire tidal tail and the entire fluid-grid.   Colors indicate temperature.  Also included are three density contours at $10^{11}$g cm${}^{-3}$ (white), $10^{10}$g cm${}^{-3}$ (red), and $10^9$g cm${}^{-3} (black)$.  The edge of the fluid grid at this time can be identified at the interface between light brown and grey.  The black hole horizon is a black circle near the middle of the $10^{11}$g cm${}^{-3}$ contour.
}
\label{fig:2dsnapshot}
\end{figure*}

\begin{figure}
\includegraphics[width=1.25\columnwidth]{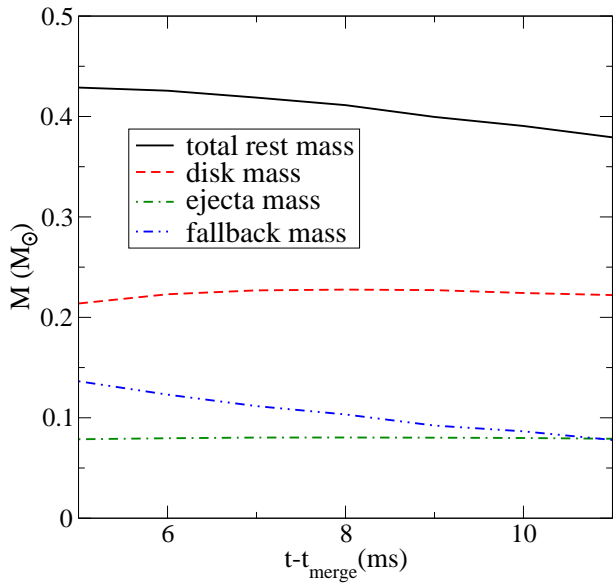}
\caption{Post-merger evolution in time for the FSU2.1 $M_{\rm NS}=1.2M_{\odot}$ case of the baryonic mass of each component of matter outside the black hole:  the disk, ejecta, and fallback.  Also plotted in solid black is their sum, the total baryonic mass outside the black hole.  The disk is depleted by accretion but replenished by fallback; its mass at first increases, then decreases.}
\label{fig:components}
\end{figure}

After the initial shock and disk formation, the remaining bound matter (specific orbital energy $e<0$) outside the black hole can be divided into two classes:  the incipient disk (what we have been calling the ``proto-disk'') and the fallback material.  It turns out to be possible to make this division fairly precise, as we see a sharp division in temperature between the inner quasi-circularized material and the outer infalling material.  (See Figure~\ref{fig:2dsnapshot}.)  Therefore, we define disk material to be bound matter with temperature above 0.2 MeV, and fallback material to be bound matter with temperature below this, and the component masses are insensitive to the choice of cutoff temperature within the range $\sim$ 0.1--1\,MeV.

The component masses are plotted as a function of time in Figure~\ref{fig:components} for one representative case.  As matter passes through the fallback-disk interface shock, it heats and circularizes, becoming part of the proto-disk.  Thus, the proto-disk is depleted by accretion into the black hole, but grows by the infusion of fallback material.  The initial fallback rate is quite high ($\approx 2 M_{\odot}$s${}^{-1}$), so that the disk initially gains mass before peaking around 8\,ms after merger, after which time accretion becomes dominant.  In our simulations, accretion is driven by hydrodynamic processes such as angular momentum transport by nonaxisymmetric disturbances.  In reality, one would expect magnetorotational effects to drive the accretion during the subsequent evolution.

\begin{figure}
\includegraphics[width=1.2\columnwidth]{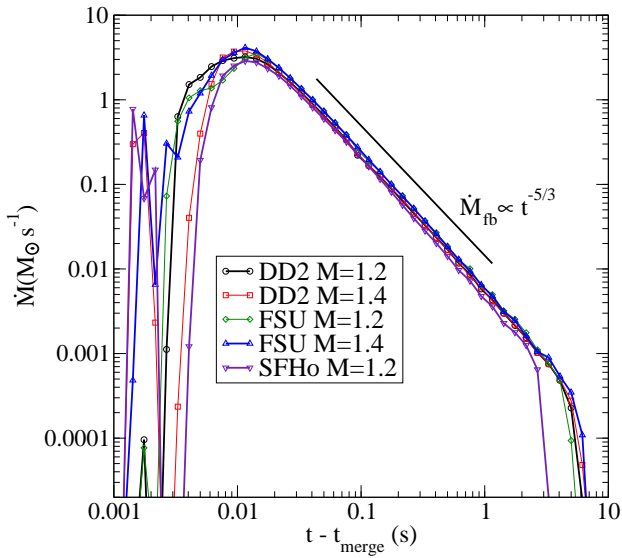}
\caption{Predicted fallback rate based on the orbital period of weakly bound material measured at 5ms after merger.  A $t^{-5/3}$ power law is included for comparison.
}
\label{fig:disk_fallback}
\end{figure}

The time it will take for the remaining fallback material to incorporate itself into the disk can be estimated from the material's Keplerian orbital period.  From the mass of material with each fallback time, a fallback rate can be calculated.  This is plotted for all cases in Figure~\ref{fig:disk_fallback}.  The fallback rate follows a $t^{-5/3}$, in agreement with expectations from the literature~\cite{Rosswog:2006rh,Chawla:2010sw}.

\begin{figure}
\includegraphics[width=1.2\columnwidth]{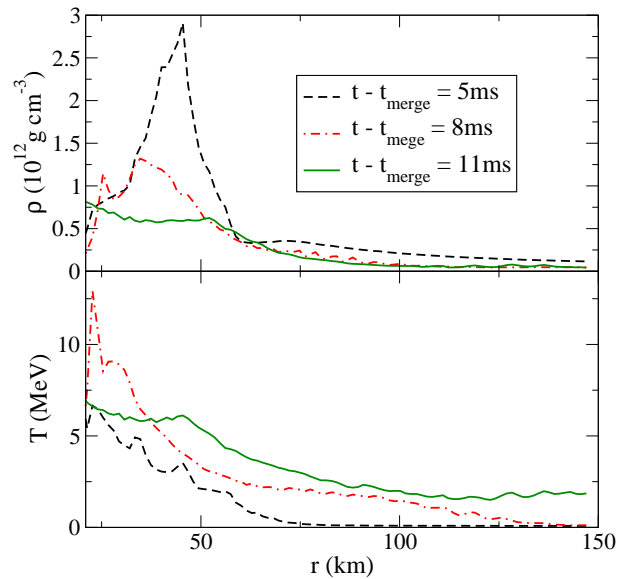}
\caption{Density-weighted averaged density and temperature of the disk in the early post-merger phase for the $M=1.2 M_{\odot}$ case with FSU2.1 EOS.
}
\label{fig:disk_fsu}
\end{figure}

\begin{figure}
\includegraphics[width=1.2\columnwidth]{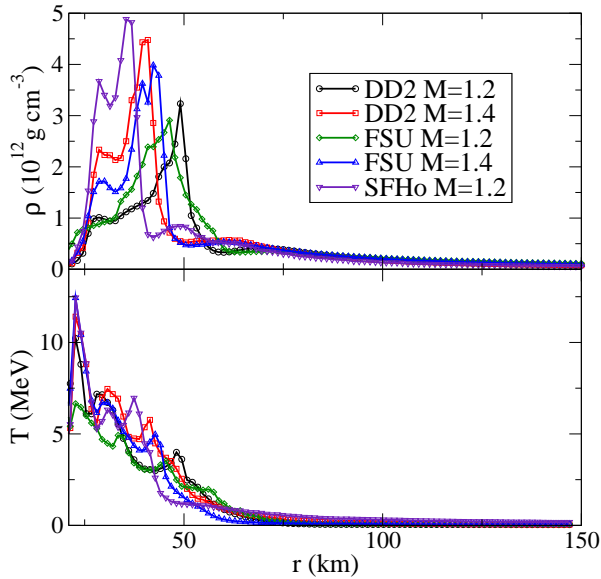}
\caption{Density-weighted averaged density and temperature profiles of the protodisks 5 ms after merger.  The densest zone in the disks ranges $R \approx 30 - 50 {\rm km}$ from the center of the black hole (with a predicted final ISCO radius of $\approx 24 {\rm km}$).  The temperature near the peaks of the density profiles is $T \sim (4-7) {\rm MeV}$.  Beyond the disk, in the fallback region, material has not yet shocked and remains quite cool.}
\label{fig:disk_5ms}
\end{figure}

\begin{figure}
\includegraphics[width=1.2\columnwidth]{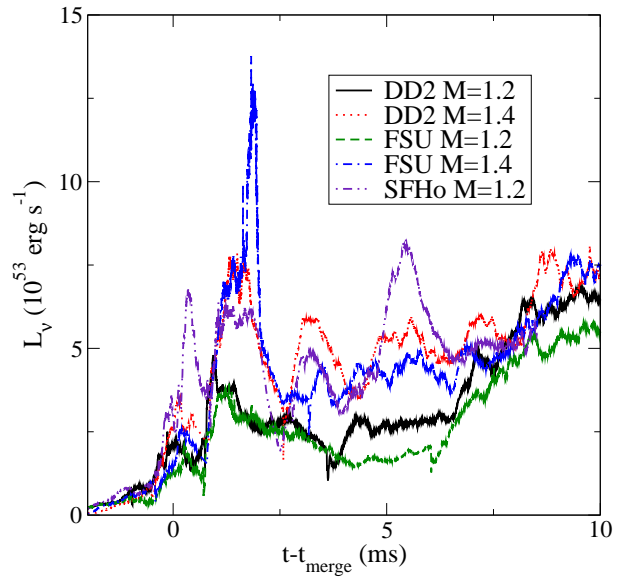}
\caption{Total neutrino luminosity as a function of time for all cases.}
\label{fig:Lnu}
\end{figure}

Radial profiles of the proto-disk for this same 1.2\,$M_{\odot}$ FSU case at various early times are plotted in Figure~\ref{fig:disk_fsu}.  A comparison of proto-disk profiles for all cases 5\,ms after merger is shown in~\ref{fig:disk_5ms}.  Each point on the radial plot represents a density-weighted average over angles.  Over the 6\,ms shown, the density and temperature profiles flatten, and the interface between shocked and unshocked material moves outward.  Neutrino transport effects omitted in this study will most likely also work to flatten the temperature profile.  We note a clear trend in the location of the early-time maximum of the density:  it tends to be closer to the black hole for more compact progenitor neutron stars.  This would affect disk properties such as the dynamical/orbital timescale, but the trend is quickly washed out as the disk profiles flatten. The resulting neutrino luminosity as a function of time is shown in Figure~\ref{fig:Lnu}.  Consistent with~\cite{Kyutoku:2017voj}, we see that the more compact stars tend to produce slightly more neutrino bright disks; even though these disks can be less massive, they can be denser and hotter.

\section{Discussion}
\label{sec:discussion}

Evolution on the accretion timescale ($\sim 100$\,ms) will be dominated by the (presumably magnetic) angular momentum transport mechanism.  These first tens of ms, however, are a distinct phase of the post-merger evolution (fallback/shock rather than MHD dominated) of interest beyond its role of constructing the subsequent ``standard'' accretion torus scenario.  A significant fraction of the total neutrino energy output may well come from this early phase, during which time the luminosity can be reasonably modeled without transport processes.  In our recent magnetohydrodynamic simulations of BHNS post-merger disks~\cite{Nouri:2017fvh}, we compared the evolution of the disk with and without a strong seed magnetic field.  As expected, the magnetic field drives long-term accretion not present in its absence.  However, the neutrino luminosity $L_{\nu}$ remains quite similar in both cases for the first $\sim$30\,ms, after which $L_{\nu}$ has dropped by an order of magnitude.  Before this drop off, shock heating from fallback accretion and disk settling maintains $L_{\nu}$ in both cases, while magnetoturbulent heating and advective cooling roughly cancel.  Viscous hydrodynamic simulations with viscosity parameter as high as $\alpha=0.1$ also find transport effects to be unimportant for the early-time energy release~\cite{Hadaddi:inprep}.

Fallback accretion is important to the disk mass and thermal energy budget at early times but not late times.  In the absence of a disk wind, a radiatively inefficient, advective disk (as the torus will quickly become) follows $\dot{M}\propto t^{-4/3}$~\cite{Metzger:2008av}, which soon dominates the fallback's steeper $\dot{M}_{\rm fb}\propto t^{-5/3}$.  Disk winds can steepen the accretion rate to $t^{-8/3}$, while numerical simulations find $\dot{M}\propto t^{-2.2}$~\cite{Fernandez:2014b}.  However, the same simulations find that the wind stops the fallback accretion after 100\,ms.

Radiative hydrodynamic evolutions suggest that BHNS disks can produce GRB fireballs by $\nu\overline{\nu}$ annihilation~\cite{Just:2015dba} (unlike NSNS mergers, where the polar outflow introduces too much baryon loading), but the energies and durations are too low to explain most short GRBs.  After the disk becomes radiatively inefficient, relativistic outflows are still possible but must be driven by magnetohydrodynamic processes such as the Blandford-Znajek effect~\cite{1977MNRAS.179..433B}.
High resolution MHD BHNS simulations find it will likely take 30\,ms or longer for such a magnetic jet to form~\cite{Kiuchi:2015qua}, so the character of the relativistic outflow might then change from fireball to Poynting flux dominated (cf.~\cite{Barkov:2011jf}).

We had hoped to identify new EOS-dependent observables, in particular something that would differentiate EOS with the same compaction.  Since we have sampled a few EOS rather than working with an EOS family with free parameters (such a thing not being available for $T$,$Y_e$-dependent EOS until very recently~\cite{PhysRevC.96.065802}), we could not do such a search systematically, e.g. by fixing compaction and varying some independent variable.  However, our low-mass soft EOS has compaction similar to our high-mass stiff EOS.  For the most part, our discussion (like most in the literature) has concentrated on differences in the cold, beta-equilibrium EOS.  Our simulations would also be sensitive to differences in the $T$ or $Y_e$ dependence (at least, those that manifest themselves below 10MeV), although in fact the thermal contributions to internal energy and pressure are quite similar for our chosen EOS, and we saw no differences in merger results that required invoking these other dimensions of the EOS.

Our results suggest that more compact neutron stars produce more compact, initially brighter, accretion disks.  We confirm dependencies of ejecta on compaction quantified in earlier works. 
However, in this large disk mass regime, the disk mass appears to be less sensitive to compaction than expected.  The use of more general EOS has not uncovered any new merger properties that seem able to provide additional EOS information.  A more systematic study would still be useful to uncover subtle EOS signatures, although the more subtle they are, the less observationally useful.

\acknowledgments
The authors thank Roland Haas and the members of the SXS collaboration
for helpful discussions over the course of this project. 
M.D. acknowledges support through NSF Grant PHY-1402916. 
Support for this work was provided by NASA through Einstein Postdoctoral Fellowship grant numbered PF4-150122 (F.F.)
awarded by the Chandra X-ray Center, which is operated by the Smithsonian Astrophysical Observatory for NASA under contract NAS8-03060.
H.P. gratefully acknowledges support from the NSERC Canada, the Canada Research Chairs Program and the Canadian Institute for Advanced Research.
L.K. acknowledges support from NSF grants PHY-1708212 and PHY-1708213 at Cornell,
while the authors at Caltech acknowledge support from NSF Grants PHY-1404569, and NSF-1440083.
Authors at both Cornell and Caltech also thank the Sherman Fairchild Foundation for their support.
Computations were performed on the supercomputer Briar\'ee from the Universit\'e de Montr\'eal,
managed by Calcul Qu\'ebec and Compute Canada. The operation of these supercomputers is funded
by the Canada Foundation for Innovation (CFI), NanoQu\'ebec, RMGA and the Fonds de recherche du Qu\'ebec - Nature et
Technologie (FRQ-NT). Computations were also performed on the Zwicky and Wheeler clusters at Caltech, supported by the Sherman
Fairchild Foundation and by NSF award PHY-0960291.

\bibliography{References}

\end{document}